\documentstyle[12pt]{article}

\def\thefootnote{\fnsymbol{footnote}}
\def\bea{\begin{eqnarray}}
\def\eea{\end{eqnarray}}
\def\beq{\begin{equation}}
\def\eeq{\end{equation}}

\def\ibar{\bar{\imath}}

\def\[{\left [}
\def\]{\right ]}
\def\({\left (}
\def\){\right )}

\def\pp{\partial}
\def\M{\bar{M}}

\def\z{\bar{z}}

\def\STr{{\rm STr}}
\def\Tr{{\rm Tr}}

\def\L{{\cal L}}
\def\D{{\cal D}}
\def\notD{\not{\hspace{-.05in}\D}}
\def\bl{\bar{\lambda}}

\def\hz{\hat{z}}

\def\hA{\hat{A}}

\def\W{\overline{W}}

\def\n{\bar{n}}
\def\m{\bar{m}}

\def\bc{\bar{\chi}}

\def\bc{\bar{\chi}}

\def\Z{{\bar{Z}}}

\def\bv{\bar{\varphi}}

\def\bj{\bar{\jmath}}

\begin{document}

\begin{titlepage}
\begin{center}

\hfill LBL-36548 \\
\hfill UCB-PTH-94/36 \\
\hfill December 1994 \\
\hfill hep-th/9412125 \\[.3in]

{\large \bf PAULI-VILLARS REGULARIZATION OF GLOBALLY SUPERSYMMETRIC
THEORIES}\footnote{This
work was supported in part by
the Director, Office of Energy Research, Office of High Energy and Nuclear
Physics, Division of High Energy Physics of the U.S. Department of Energy under
Contract DE-AC03-76SF00098 and in part by the National Science Foundation under
grant PHY--90--21139.} \\[.2in]

Mary K. Gaillard \\[.1in]

{\em Department of Physics and Theoretical Physics Group,
 Lawrence Berkeley Laboratory,
 University of California, Berkeley, California 94720}\\[.5in]

\end{center}

\begin{abstract}

It is shown that the one-loop ultraviolet divergences in
renormalizable supersymmetric theories can be regulated by
the introduction of heavy Pauli-Villars chiral supermultiplets,
provided the generators of the gauge group are traceless in the matter
representation.  The procedure is extended to include supersymmetric gauged
nonlinear sigma models.

\end{abstract}
\end{titlepage}

\newpage
\renewcommand{\thepage}{\arabic{page}}
\setcounter{page}{1}
\def\thefootnote{\arabic{footnote}}
\setcounter{footnote}{0}

It was recently shown~\cite{mk} that the one-loop quadratic divergences of
standard supergravity
can be regulated by the introduction of heavy Pauli-Villars fields belonging to
chiral and abelian gauge multiplets.  This result holds {\it a fortiori} for
renormalizable supersymmetric theories.  While it has not yet been shown that
the full supergravity theory, including all logarithmic divergences, can be
regulated in this way, the results of supergravity
calculations~\cite{us},~\cite{us2}
can be used to show that this is true for the easier case of renormalizable
supersymmetric theories, provided Tr$(T_a)$ vanishes, where the $T_a$ represent
the generators of the gauge group on the light chiral multiplets.
This result may be of some practical value in calculating
quantum corrections, although there is no objection of principle to using
dimensional regularization for these calculations.  I also consider the class
of nonrenormalizable theories with a nontrivial K\"ahler potential; in this
case
the Pauli-Villars masses, which play the role of effective cut-offs, acquire
physical significance.

The Pauli-Villars regularization procedure is presented
here using functional integration in an arbitrary bosonic background.  That is,
I calculate the one-loop correction to that part of the bosonic action that
grows with the effective cut-off(s).  Since the result just amounts to wave
function renormalizations (and/or renormalization of the K\"ahler potential),
the fermionic action can be inferred from supersymmetry.
The derivative expansion used here to obtain the ultraviolet divergent part of
the effective action is not generally amenable to the calculation of finite
S-matrix matrix elements because higher order terms in the expansion are
infrared divergent.  However, since it is shown that the theory is
ultraviolet finite (at least at one loop) with the appropriate
choice of Pauli-Villars fields, there is no impediment in principle to
implementing this regularization procedure in Feynman diagram calculations.

The one-loop effective action $S_1$ is obtained from the term quadratic in
quantum fields when the Lagrangian is expanded about an arbitrary background;
for a renormalizable gauge theory:
\bea \L_{quad}(\Phi,\Theta,c) &=& -{1\over 2}\Phi^TZ^\Phi\(\D^2_\Phi +
H_\Phi\)
\Phi + {1\over 2}\bar{\Theta}Z^\Theta\(i\notD_\Theta - M_\Theta\)\Theta
\nonumber \\ & & + {1\over 2}\bar{c} Z^c\(\D^2_c + H_c\)c + O(\psi), \eea
where $\D_\mu$ is the gauge covariant derivative,
the column vectors $\Phi,\Theta,c$ represent quantum bosons, fermions
and ghost fields, respectively, and $\psi$ represents background fermions that
are set to zero throughout this paper.  The one loop bosonic action is
given by
\bea S_1 &=& {i\over 2}\Tr\ln\(\D_\Phi^2 + H_\Phi\)
-{i\over 2}\Tr\ln\(-i\notD_\Theta + M_\Theta\)
+ {i\over2}\STr\ln\(\D_c^2 + H_c\)
\nonumber \\ &=& {i\over 2}\STr\ln\(\D^2 + H\) + T_-, \eea
where $T_-$ is the helicity-odd fermion contribution which is finite for a
renormalizable theory (and, more generally, in the absence of a
dilaton~\cite{us2}), and the helicity-even contribution is determined by
\beq \D^2_\Theta + H_\Theta \equiv
\(-i\notD_\Theta + M_\Theta\)\(i\notD_\Theta + M_\Theta\).\eeq
The field-dependent matrices $H(\phi)$ and
$\D_\mu(\phi)$, as extracted by taking the flat (K\"ahler and space-time)
limit of the results of~\cite{us},~\cite{us2}, are given below.
Explicitly evaluating (2) with an ultraviolet cut-off $\Lambda$ and a
massive Pauli-Villars chiral supermultiplet sector with a squared mass matrix
of the form
$$M_{PV}^2 = H^{PV}(\phi) + \pmatrix{\mu^2& \nu\cr \nu^{\dag}&\mu^2\cr} \equiv
H^{PV} + \mu^2 + \nu, \;\;\;\; |\nu|^2\sim \mu^2\gg H^{PV}\sim H, $$
gives, with $H' = H + H^{PV}$:
\bea 32\pi^2S_1 &=& - \int d^4xd^4p\STr\ln\(p^2 + \mu^2 + H' + \nu\)
+ 32\pi^2\(S'_1 + T_-\)
\nonumber \\
&=& 32\pi^2\(S'_1 + T_-\) - \int d^4xd^4p\STr\ln\(p^2 + \mu^2\) \nonumber \\ &
&
- \int d^4xd^4p\STr\ln\[1 + \(p^2 + \mu^2\)^{-1}\(H' + \nu\)\] , \eea
where $S'_1$ is a logarithmically divergent contribution that involves the
Yang-Mills field strength, $G_{\mu\nu} = [\D_\mu,\D_\nu]$:
\beq 32\pi^2 S'_1 = {1\over12}\int d^4xd^4p\STr{1\over\(p^2 + \mu^2\)}
G'_{\mu\nu}{1\over\(p^2 + \mu^2\)}G'^{\mu\nu}, \;\;\;\; G'_{\mu\nu} =
G_{\mu\nu} + G^{PV}_{\mu\nu}. \eeq

Finiteness of (4) requires
\beq  \STr \mu^{2n} = \STr H' = \STr\(2\mu^2H' + \nu^2\) =
\STr H'^2 + {1\over6}\STr G'^2 = 0.\eeq
The vanishing of $\STr \mu^{2n}$ is automatically assured by supersymmetry.
Once the remaining conditions are satisfied we obtain
\beq S_1 = - \int {d^4x\over64\pi^2}\STr\[\(2\mu^2 H' + \nu^2 + H'^2
 + {1\over6}G'_{\mu\nu}G'^{\mu\nu}\)\ln\mu^2\] + {\rm finite\;terms}.\eeq

The Lagrangian for light fields is
\bea
\L &=& \D^\mu z^i\D_\mu\z_i - {1\over 4}F_{\mu\nu}F^{\mu\nu}
- W_i\W^i - {g^2\over2}\(\z T^az\)\(\z T_az\) \nonumber \\
& & + {i\over 2}\(\bl\notD\lambda + \bc^L_i\notD\chi_L^i +
\bc_R^i\notD\chi^R_i\) \nonumber \\ & &
+ \sqrt{2}g\(i\bl^a_R(T_a\z)_i\chi^i_L + {\rm h.c.}\) + \L_{gf} + \L_{gh},
\eea
where $Z^i = (z^i,\chi^i_L)$ is a chiral supermultiplet, $\lambda^a$ is a
gaugino, $F^a_{\mu\nu}$ is the Yang-Mills field strength, and
$W(z)$ is the superpotential.  The gauge-fixing and ghost terms~\cite{us} are,
respectively,
\bea
\L_{gf} &=& - {1\over 2}C_aC^a, \;\;\;\; C^a = \D^\mu\hA^a_\mu + i
g\[(\z T^a\hz) - ({\hat{\bar{z}}}T^a\z)\], \nonumber \\
\L_{gh} &=& {1\over 2}\bar{c}_a\[\(\D_\mu\D^\mu\)^a_b + g^2(\z T^az)(\z T_bz)
+ g^2(\z T_bz)(\z T^az)\]c^b, \eea
where quantum variables are hatted and background fields are unhatted.

To regulate the theory we introduce Pauli-Villars regulator chiral
supermultiplets $\phi_{PV}$:
$Z_\alpha^I = (\Z^\alpha_{\bar{I}})^{\dag}$, $Z'^I_\alpha =
(\Z'^\alpha_{\bar{I}})^{\dag}$ and $\varphi^a_\beta = (\bv_a^\beta)^{\dag}$,
with signature $\eta^{\alpha,\beta} = \pm 1$, which determines the
sign of the corresponding contribution to the supertrace relative to an
ordinary
particle of the same spin.  Thus $\eta = +1 (-1)$ for ordinary particles
(ghosts). $Z^I$ transforms like $z^i$ under the gauge group,
$Z'^I$ transforms according to the representation conjugate to $z^i$, and
$\varphi^a$ transforms according to the adjoint representation. Including
these fields the superpotential is
\bea W(Z^i,Z^I,Z'^I,\varphi^a) &=& W(Z) +
\sum_{\alpha,I}\mu^\alpha_IZ^I_\alpha Z'^I_\alpha
+ \sum_{a,\beta}\mu_\beta\varphi_a^\beta\varphi^a_\beta \nonumber \\ & &
+ {1\over 2}\sum_\alpha a^\alpha W_{ij}Z^I_\alpha Z^J_\alpha +
g^2\sum_\gamma b^\gamma\theta^a_\gamma(Z'_\gamma T_aZ), \eea
where the ranges of summation are
\beq
\alpha = 1,\cdots, N_I, \;\;\;\; \beta = 1,\cdots, N_\varphi, \;\;\;\;
\gamma = 1,\cdots, N_{I\varphi}, \;\;\;\; N_{I\varphi} \le {\rm min}\{
N_I,N_\varphi\}, \eeq
The parameters $\mu$ play the role of effective cut-offs, and $a^\alpha,
b^\gamma$ are of order unity.

The supertraces needed to evaluate the divergent part of the one-loop bosonic
action are~\cite{us}--\cite{josh}
\bea \STr H_\chi &=&  2g^2(\z T_az)\Tr(T^a), \;\;\;\; \STr H_g = 0,\eea
where the subscripts $\chi,g$ refer to supertraces over chiral and Yang-Mills
(including ghosts) supermultiplets, respectively.  Since the heavy
Pauli-Villars fields form vector-like representations, they do not contribute
to Tr$(T_a)$; thus we require Tr$(T_a) = 0$ for the light fields.

In addition, defining
\beq X_AY^A = \sum_A \eta^A X_AY^A \eeq
we have, for traces evaluated with $\phi_{PV}=0$:
\bea {1\over 2}\Tr H_\chi^2 &=& W_{iAB}\W^{ABj}\(\W^iW_j
+ \D_\mu z^i\D^\mu\z_j\)
\nonumber \\ & & - g^2(\z T^az)\[(T_az)^iW_{iAB}\W^{AB}
- 4g^2\(\z C_2(R_z)T_az\)\]  \nonumber \\ & & - 2g^4C_G^a(\z T_az)(\z T^az)
- {g^2\over2}C^a_\chi\[F^{\mu\nu}_aF_{\mu\nu}^a - 2 g^2
(\z T_az)(\z T^az)\], \nonumber \\
{1\over 2}\STr H_{\chi g}^2 &=& - 4g^2\(\D_\mu\z C_2(R_z)T_a\D^\mu z\)
- 4g^2(T_az)^i(T^a\z)^{\n}W_{ki}\W_{\n}^k \nonumber \\
{1\over 2}H^2_g &=&  g^2C_G^a\[{3\over2}F^a_{\mu\nu}F_a^{\mu\nu}
- g^2(\z T_az)(\z T^az)\], \nonumber \\
\STr G^\chi_{\mu\nu}G_\chi^{\mu\nu} &=&
\STr G^{g}_{\mu\nu}G_{g}^{\mu\nu} = 0.
\eea
In these expressions, $A,B$ refer to all (light and Pauli-Villars) chiral
multiplets, $C_G^a$ is the Casimir in the adjoint representation of the gauge
subgroup $G_a$, $C_2(R)$ is the Casimir in the chiral multiplet
representation $R$: $\(\z T^aT_az\) = \(\z C_2(R_z)z\)$,
and $C^a_\chi$ is defined by
\bea C^a_\chi\delta_{ab} &=& [\Tr(T_aT_b)]_{{\rm light+PV}},
\;\;\;\; C^a_\chi = \sum_{R_i}C^a_{R_i}\(1 + 2\sum_\alpha\eta^\alpha_I\)
+ C^a_G\sum_\beta\eta_a^\beta,\;\;\;\; \nonumber \\
C^a_R\delta_{ab} &=& [\Tr(T_aT_b)]_R ,\;\;\;\; C^a_R =
{{\rm dim}(R)\over
{\rm dim}(G_a)}C^a_2(R), \;\;\;\; C_2(R) = \sum_aC^a_2(R).\;\;\;\;\eea
We obtain for the overall supertraces:
\bea {1\over 2}\STr H &=&
\[W_{iAB}\W^{ABj} - 4g^2\delta_i^jC_2(R_i)\]
\(\W^iW_j + \D_\mu z^i\D^\mu\z_j\)
\nonumber \\ & & - g^2(\z T^az)\[(T_az)^iW_{iAB}\W^{AB}
- 4g^2\(\z C_2(R_z)T_az\)\] \nonumber \\ & &
+ {g^2\over2}\(3C^a_G - C^a_\chi\)\[F^{\mu\nu}_aF_{\mu\nu}^a - 2 g^2
(\z T_az)(\z T^az)\], \nonumber \\
\STr G_{\mu\nu}G^{\mu\nu} &=& 0,
\eea
where I used the identities
\beq (T^az)^iW_i = 0 = \pp_j\[(T^az)^iW_i\] = (T^a)^i_jW_j + (T^az)^iW_{ij},
\eeq that follow from the gauge invariance of the superpotential.
Explicitly evaluating the derivatives of the superpotential with respect to
the Pauli-Villars fields gives:
\bea (T_az)^iW_{iAB}\W^{AB} &=& (T_az)^iW_{ijk}\W^{jk}\(1 +
\sum_\alpha\eta^\alpha_Ia^2_\alpha\)
+ g^2\(\z C_2(R_z)T_az\)\sum_\gamma\eta^\gamma b^2_\gamma,  \nonumber \\
W_{iAB}\W^{ABj} &=& W_{ik\ell}\W^{jk\ell}\(1 +
\sum_\alpha\eta^\alpha_Ia^2_\alpha\) +
g^2\delta^j_iC_2(R_i)\sum_\gamma\eta^\gamma b^2_\gamma. \eea
Therefore finiteness requires:
\beq 0 = 1 + 2\sum_\alpha\eta^\alpha_I = 3 - \sum_\gamma\eta^\gamma_a = 1 +
\sum_\alpha\eta^\alpha_Ia^2_\alpha = 4 - \sum_\gamma\eta^\gamma_a b^2_\gamma.
\eeq

To extract the residual finite part, we may take, for example, \bea
\eta^\alpha_I = \eta^\alpha, \;\;\;\;\mu^\alpha_I &=& \mu\delta^\alpha,
\;\;\;\; \mu^\beta = \mu\epsilon^\beta, \eea
and use the results of~\cite{sigma} to evaluate the terms in (7) that grow with
the effective cut-off $\mu$.  Since there are no terms of order $\mu^2$ in
(14), we need only:
\beq \sum_\alpha\eta_\alpha\ln(\mu^2\lambda_\alpha) =
\ln\mu^2\sum_\alpha\eta^\alpha + \ln\rho, \;\;\;\;
\ln\rho = \sum_\alpha\eta^\alpha\ln\lambda_\alpha.\eeq
Then we obtain for the one-loop effective action:
\bea \L_1 &=& - {1\over64\pi^2}\STr\(H'^2\ln\mu^2\) +
{\rm finite\;terms} \nonumber \\ &=& {1\over 64\pi^2}\Bigg\{
\[\ln(\mu^2/\rho'_Zm'^2_Z)W_{ik\ell}\W^{jk\ell} - 4g^2\delta_i^jC_2(R_i)
\ln(\mu^2\rho_{Z\varphi}/m^2_{Z\varphi})\]\(\W^iW_j + \D_\mu z^i\D^\mu\z_j\)
\nonumber \\ & & - g^2(\z T^az)\[\ln(\mu^2/\rho'_Zm'^2_z)(T_az)^iW_{ijk}\W^{jk}
- 4g^2\ln(\mu^2/\rho_{Z\varphi}m^2_{Z\varphi})\(\z C_2(R_z)T_az\)\]
\nonumber \\ & &
+ {g^2\over2}\[3C^a_G\ln(\mu^2\rho_\varphi/m^2_\varphi) -
C^a_M\ln(\mu^2/\rho'_Zm'^2_Z)\]
\[F^{\mu\nu}_aF_{\mu\nu}^a - 2 g^2(\z T_az)(\z T^az)\]\Bigg\}
\nonumber \\ & & + {\rm finite\;terms} .\eea
The $m$'s are the appropriate infrared cut-offs, and
\bea \ln\rho_Z &=& 2\sum_\alpha\eta^\alpha\ln\delta^2_\alpha, \;\;\;\;
\ln\rho_\varphi = \sum_\beta\eta^\beta\ln\epsilon^2_\beta, \nonumber \\
\ln\rho'_Z &=& \sum_\alpha\eta^\alpha\ln(\delta_\alpha a_\alpha)^2, \;\;\;\;
\ln\rho_{Z\varphi} = \sum_\gamma\eta^\gamma
\ln(\delta_\gamma\epsilon_\gamma b_\gamma^2).
\eea
Then the one-loop corrected bosonic Lagrangian can be written as
\beq \L_0(z,A_\mu,g) + \L_1 = \L_0\(Z_2^{-{1\over2}}z,Z_3^{-{1\over2}}A_\mu,
Z_3^{1\over2}g\) + {\rm finite\;terms},\eeq
with \bea
\(Z^{-1}_2\)^j_i &=& {1\over32\pi^2}\[\ln(\mu^2/\rho'_Zm'^2_Z)
W_{ik\ell}\W^{jk\ell} - 4g^2\delta^j_i\ln\(\mu^2\rho_{Z\varphi}/m^2_{Z\varphi}
\)\sum_aC_2^a(R_i)\],\;\;\;\; \nonumber \\
Z^a_3 &=& {g^2\over 16\pi^2}\[3C^a_G\ln(\mu^2\rho_\varphi/m^2_\varphi) -
C^a_M\ln(\mu^2/\rho'_Zm'^2_Z)\], \;\;\;\; C^a_M = \sum_RC^a_R.\;\;\;\;\eea
The coefficients of ln$\mu^2$ in the renormalization constants correspond to
the
standard result for the SUSY $\beta$-function and to the chiral wave function
renormalization found in ``supersymmetric'' gauges~\cite{barb},~\cite{us} and
in
string loop calculations~\cite{ant}.

It is straightforward to extend these results to the nonlinear $\sigma$-model,
again using the results of~\cite{us},~\cite{us2} for supergravity. We introduce
an arbitrary K\"ahler potential $K(Z^i,\Z^i)$, with K\"ahler metric $K_{i\m}$;
as in~\cite{mk}, $Z^I,Z'^I$ have the same K\"ahler metric as $Z^i$, while
$\varphi^a$ has K\"ahler metric $\delta_{ab}$. However to regulate the
logarithmic divergences, we have to specify additional terms in $K$.  With the
conventions that indices are raised and lowered with the K\"ahler metric, and
that scalar derivatives are field reparameterization covariant:
\bea \D_\mu z^i\D^\mu\z_i &=& \D_\mu z^i\D^\mu\z^{\m}K_{i\m},\;\;\;\; \Z_I =
K_{i\m}(z^j,\z^{\bj})\Z^{\M}, \nonumber \\
W_{ij} &=& \pp_iW_j + \Gamma^k_{ij}W_k \;\;\;\; {\rm etc.}, \eea
we take
\bea & & K(Z^i,\Z^{\ibar},\phi_{PV}) = K(Z^i,\Z^{\ibar}) + \sum_\alpha
\(Z_\alpha^I\Z^\alpha_I + Z'^I_\alpha\Z'^\alpha_I\) +
\sum_\beta \bv^\beta_a\varphi^a_\beta \;\;\;\;\;\;\;\;\;\nonumber \\ & & \qquad
+ {1\over2}\[\sum_{\alpha;\;I,J=i,j}K_{ij}(z^k,\z^{\bar{k}})
\(Z_\alpha^IZ^J_\alpha + Z'^I_\alpha Z'^J_\alpha\) + {\rm h.c.}\] +
O\(\phi_{PV}^2\),\;\;\;\;\; \eea
and, Eqs. (12) and (14) are modified as follows: \bea
\STr H_\chi &=& 2g^2\D^aD_A(T_az)^A - 2R_i^j\(\W^iW_j
+ \D_\nu z^i\D^\mu\z_j\),\nonumber \\
D_A(T_az)^A &=& \Tr(T_a) + \Gamma^A_{Ai}(T_az)^i = \Tr(T_a) +
\(1 + 2\sum_\alpha\eta^\alpha_I\)\Gamma^A_{Ai}(T_az)^i, \nonumber \\
\D_a &=& K_i(T_az)^i, \;\;\;\;  R_i^j = R_{\;i\;A}^{j\;A} =
\(1 + 2\sum_\alpha\eta^\alpha_I\)R_{\;i\;k}^{j\;k}, \eea
so STr$H_\chi$ vanishes by the conditions (19) with Tr$(T_a) = 0$.  Here
$R^A_B$ and $R^{A\;\;C}_{\;\;B\;\;D}$ are the K\"ahler Ricci and Riemann
tensors, respectively.  In addition:
\bea {1\over 2}\Tr H_\chi^2 &=& \(W_{iAB}\W^{ABj} + 2W_{AB}\W^{AC}
R^{B\;j}_{\;C\;i} + W_k\W^{\ell}R^{A\;j}_{\;B\;\ell}R^{B\;k}_{\;A\;i}\)
\(\W^iW_j + \D_\mu z^i\D^\mu\z_j\) \nonumber \\ & &
+ 2\D_\mu z^j\D^\mu\z_iR^{i\;B}_{\;j\;A}R^{\ell\;A}_{\;k\;B}W_{\ell}\W^k
\nonumber \\ & & + \D_\mu z^j\D^\mu z^i\(W_{iAB}\W^kR^{A\;\;B}_{\;\;k\;\;j}
- R^{A\;B}_{\;\;j\;\;i}W_{kAB}\W^k +{\rm h.c.}\) \nonumber \\ & &
 + \D_\mu z^j\D^\mu\z_i\D_\nu z^{\ell}\D^\nu\z_k\(R^{A\;i}_{\;B\;j}
R^{B\;\;k}_{\;\;A\;\;\ell} -
{1\over2}R^{A\;i}_{\;B\;\ell}R^{B\;\;k}_{\;\;A\;\;i}\) \nonumber \\ & &
+ \D_\mu z^j\D^\mu z^i\D_\nu\z_k\D^\nu\z_{\ell}\(R^{A\;B}_{\;\;j\;\;i}
R^{k\;\ell}_{\;A\;B} + {1\over2}R^{A\;k}_{\;B\;i}
R_{\;B\;\ell}^{A\;j}\)
\nonumber \\ & & - g^2\D^a(T_az)^iD_i\[W_{AB}\W^{AB} - 4g^2
(\z T^b)_j(T_bz)^j\]  - 2g^4C_G^a\D_a\D^a \nonumber \\ & &
- {g^2\over2}D_A(T^bz)^BD_B(T_az)^A\(F^{\mu\nu}_bF_{\mu\nu}^a - 2 g^2
\D_a\D^a\) \nonumber \\ & &
+ 4g^2(\z T_b)_{\ell}(T^bz)^k\[R^{\ell\;i}_{\;k\;j}W_i\W^j +
2g^2\D_aR^{A\;j}_{\;B\;i}W_j\W^iD_A(T^az)^B\] \nonumber \\ & &
+ 2g^2\D_\mu z^i\D^\mu\z_j\[2 R^{j\;k}_{\;i\;\ell}(\z T^a)_k(T_az)^{\ell}
+ \D_a R^{j\;A}_{\;i\;B}D_A(T^az)^B\] \nonumber \\ & &
+ g^2\D_a(T^az)^iR_{i\;\;j}^{\;A\;\;B}\W^jW_{AB}
- 2iF^a_{\mu\nu}D_A(T_az)^BR^{A\;\;j}_{\;\;B\;\;i}\D^\mu z^i\D^\nu\z_j,
\nonumber \\
{1\over 2}\STr H_{\chi g}^2 &=& - 4g^2D^k(T_a\z)_i\D_\mu\z_k
D_j(T^az)^i\D^\mu z^j - 4g^2(T_az)^i(T^a\z)^{\n}W_{ki}\W_{\n}^k \nonumber \\
{1\over 2}H^2_{g} &=&  g^2C_G^a\({3\over2}F^a_{\mu\nu}F_a^{\mu\nu}
- g^2\D_a\D^a\). \eea
Using the conditions (19) we have
\bea D_A(T^bz)^BD_B(T_az)^A &=& D_i(T^bz)^jD_j(T_az)^i\(1 +
2\sum_\alpha\eta^\alpha_I\) + \delta^b_aC^a_G\sum_\beta\eta^\beta_a
= 3\delta^b_aC^a_G, \nonumber \\
(T_az)^iD_iW_{AB}\W^{AB} &=& (T_az)^iW_{ijk}\W^{jk}\(1 +
\sum_\alpha\eta^\alpha_Ia^2_\alpha\) \nonumber \\ & &
+ g^2(\z T^b)(T_az)^iD_i(T_bz)^j\sum_\gamma\eta^\gamma_a b^2_\gamma +O(\mu^2)
\nonumber \\ &=& 4g^2(\z T^b)(T_az)^iD_i(T_bz)^j +O(\mu^2), \nonumber \\
W_{iAB}\W^{ABj} &=& W_{ik\ell}\W^{jk\ell}\(1 +
\sum_\alpha\eta^\alpha_Ia^2_\alpha\) \nonumber \\ & & +
g^2D_i(T_bz)^kD^j(\z T_b)_k\sum_\gamma\eta^\gamma_a b^2_\gamma +O(\mu^2)
\nonumber \\ &=& 4g^2D_i(T_bz)^kD^j(\z T_b)_k +O(\mu^2) \nonumber \\
W_{AB}\W^{AC}R^{B\;j}_{\;C\;i} &=& W_{k\ell}\W^{km}R^{\ell\;j}_{\;m\;i}
\(1 + 2\sum_\alpha\eta^\alpha_I\) \nonumber \\ & &
- g^2(\z T_az)_{\ell}(T^az)^k
R^{\ell\;j}_{\;k\;i}\sum_\gamma\eta^\gamma_a b^2_\gamma  + O(\mu^2)
\nonumber \\ &=& - 4g^2(\z T_az)_{\ell}(T^az)^kR^{\ell\;j}_{\;k\;i} + O(\mu^2),
\nonumber \\
R^{A\;j}_{\;B\;\ell}R^{B\;k}_{\;A\;i} &=& R^{m\;j}_{\;\;n\;\ell}
R^{n\;k}_{\;m\;i}\(1 + 2\sum_\alpha\eta^\alpha_I\) = 0, \nonumber \\
W_{iAB}R^{A\;\;B}_{\;\;k\;\;j} &=& W_{i\ell m}
R^{\ell\;\;m}_{\;\;k\;\;j}\(1 + \sum_\gamma\eta^\gamma a^2_\gamma\) = 0,
\nonumber \\
R^{A\;i}_{\;B\;\ell}R^{B\;\;k}_{\;\;A\;\;i} &=&
R^{j\;\;i}_{\;m\;\;\ell}R^{m\;k}_{\;\;j\;\;i}\(1 + 2\sum_\alpha\eta^\alpha_I\)
= 0,\nonumber \\
R^{A\;B}_{\;\;j\;\;i}R^{k\;\ell}_{\;A\;B} &=&
R^{k\;m}_{\;\;j\;\;i}R^{k\;\ell}_{\;k\;m}\(1 + 2\sum_\alpha\eta^\alpha_I\)
= 0,\nonumber \\
R^{A\;j}_{\;B\;i}D_A(T^az)^B &=&  R^{k\;j}_{\;\ell\;\;i}D_k(T^az)^{\ell}\(1 +
2\sum_\alpha\eta^\alpha_I\) = 0. \eea
Inserting these results in (29), we see that the overall supertrace is finite.
The $O(\mu^2)$ terms in (30) are the contributions from $Z^I,\Z'^I$ given
in~\cite{mk}, and they contribute an $O(\mu^2)$ correction to the
K\"ahler potential that was evaluated in that paper.

The residual finite terms that grow as ln$\mu^2$ can be evaluated using (21);
as shown in~\cite{us}, in the absence of gauge couplings, they can be absorbed
into a redefinition of the superpotential, except for some terms that
depend the K\"ahler Riemann tensor\footnote{The next to last line of Eq.(3.6)
of~\cite{us} should be deleted; other corrections to~\cite{us}, that do not
affect the final result presented in that equation, are given in~\cite{us2}.},
and that correspond to higher dimension
superfield operators.  When the gauge couplings are included~\cite{us2},
there is an additional correction to the K\"ahler potential, giving:
\bea \delta K(z,\z) = {1\over32\pi^2}\[\ln(\mu^2/\rho'_Zm'^2_Z)
W_{ij}\W^{ij} - 4g^2\ln\(\mu^2\rho_{Z\varphi}/m^2_{Z\varphi}\)
(\z T_a)_i(T^a z)^i\].\;\; \eea
The bosonic part of the correction to the couplings of the Yang-Mills
superfields takes the form:
\bea \delta\L_{gauge} &=& -{g^2\over 16\pi^2}\({1\over4}
F^{\mu\nu}_bF_{\mu\nu}^a - {g^2\over2}\D_a\D^a\)\Bigg[3C^a_G\delta^a_b
\ln(\mu^2\rho_\varphi/m^2_\varphi) \nonumber \\ & & - \ln(\mu^2/\rho'_Zm'^2_Z)
D_i(T^bz)^jD_j(T_az)^i\Bigg]. \eea

Aside from the holomorphic anomaly~\cite{dkl} that arises from the
field-dependence of the infrared regulator masses, the coefficient of
$F^{\mu\nu}F_{\mu\nu}$ is not a holomorphic function.  That is,
when the K\"ahler metric is not flat, there are
corrections that correspond to D-terms as well as the usual F-terms.

It was shown in~\cite{mk}, in the context of supergravity, that Pauli-Villars
regularization of the quadratic divergencies is still possible when the theory
includes a dilaton supermultiplet coupled to the Yang-Mills supermultiplet; the
regularization of all one-loop ultraviolet divergences when a dilaton is
present will be considered elsewhere.

\vskip .3in
\noindent{\bf Acknowledgements.} I thank Hitoshi Murayama for his interest in
this work.  This work was supported in part by the
Director, Office of Energy Research, Office of High Energy and Nuclear Physics,
Division of High Energy Physics of the U.S. Department of Energy under Contract
DE-WC03-76SF00098 and in part by the National Science Foundation under grant
PHY-90-21139.

\end{document}